\documentclass[preprint,12pt]{elsarticle}

\usepackage{bm}
\usepackage{amsmath}
\usepackage{graphicx}

\usepackage{amssymb}
\journal{Annals of Physics}

\begin{document}

\begin{frontmatter}
\title{On refractive processes in strong laser field quantum electrodynamics}

\author{A. Di Piazza\corref{Corresponding author}}
\ead{dipiazza@mpi-hd.mpg.de}
\address{Max-Planck-Institut f\"ur Kernphysik, Saupfercheckweg 1, 69117 Heidelberg, Germany}

\begin{abstract}
Refractive processes in strong-field QED are pure quantum processes, which involve only external photons and the background electromagnetic field. We show analytically that such processes occurring in a plane-wave field and involving external real photons are all characterized by a surprisingly modest net exchange of energy and momentum with the laser field, corresponding to a few laser photons, even in the limit of ultra-relativistic laser intensities. We obtain this result by a direct calculation of the transition matrix element of an arbitrary refractive QED process and accounting exactly for the background plane-wave field. A simple physical explanation of this modest net exchange of laser photons is provided, based on the fact that the laser field couples with the external photons only indirectly through virtual electron-positron pairs. For stronger and stronger laser fields, the pairs cover a shorter and shorter distance before they annihilate again, such that the laser can transfer to them an energy corresponding to only a few photons. These results can be relevant for future experiments aiming to test strong-field QED at present and next-generation facilities.
\end{abstract}


\end{frontmatter}

\section{Introduction}

Nonlinear processes have always played a fundamental role in different areas of physics, spanning from hydrodynamics, atomic and laser physics to plasma and high-energy physics \cite{Scott_b_2005}. From a theoretical point of view the description of such nonlinear processes, though attractive, is also particularly challenging. Since the invention of the laser, it was manifest that one of its unique features, the coherence, would allow for the experimental investigation of nonlinear phenomena. In a laser beam, in fact, a large number of photons propagate in phase and, depending on the laser intensity and on the process at hand, they may act cooperatively. One example is atomic high-order harmonic generation (HHG), in which a large number of laser photons is absorbed by a single atom and only one high-energy photon is emitted (see the reviews \cite{Agostini_2004,Midorikawa_2011}). When laser-driven electrons (mass $m$ and charge $e<0$) are bound in atoms, nonlinear phenomena start at laser field amplitudes $E_0$ of the order of the typical atomic binding field $E_{\text{at}}=m^2|e|^5$, which corresponds to a laser intensity of $I_{\text{at}}=E_{\text{at}}^2/4\pi=7.0\times 10^{16}\;\text{W/cm$^2$}$ (units with $\hbar=c=1$ are employed throughout). In this case the average number of photons absorbed from the laser by the electron is of the order of $U_p/\omega_0$, where $U_p=e^2E_0^2/m\omega_0^2$ is its ponderomotive energy and $\omega_0$ is the central laser photon energy. HHG has also been observed for free electrons driven by an intense laser beam, being named nonlinear Thomson or nonlinear Compton scattering, depending on if quantum effects are negligible or not \cite{Moore_1995,Bula_1996}. In both nonlinear Thomson and Compton scattering, the typical electric field strength, at which nonlinear effects set on, is given by $E_{\text{rel}}=m\omega_0/|e|$. The corresponding intensity is of the order of $10^{18}\;\text{W/cm$^2$}$ at optical photon energies $\omega_0\approx 1\;\text{eV}$. An electron in a laser field with central laser photon energy $\omega_0$ and electric field strength of the order of $E_{\text{rel}}$ is accelerated to relativistic velocities already within one laser period and its dynamic becomes highly nonlinear with respect to the laser field amplitude \cite{Landau_b_2_1975}. On the other hand, quantum effects such as the recoil of the photons emitted by the laser-driven electron, strongly modify the emission process when the electric field strength of the laser in the initial rest frame of the incoming electron is of the order of the so-called critical field $E_{\text{cr}}=m^2/|e|$ of QED, corresponding to the laser intensity $I_{\text{cr}}=4.6\times 10^{29}\;\text{W/cm$^2$}$ \cite{Di_Piazza_2012}. Relativistic quantum effects also allow for the nonlinear interaction of a photon with a laser field, as in the case of electron-positron pair photo-production (nonlinear Breit-Wheeler pair production (NBWPP)) \cite{Ritus_1985,Heinzl_2010,Titov_2012,Krajewska_2013}. This process, as well as any QED process occurring in the collision of a photon with a strong laser field\footnote{The expressions ``laser field'' and ``plane wave'' will be used as synonyms throughout.}, is essentially controlled by the two Lorentz- and gauge-invariant parameters $\xi=E_0/E_{\text{rel}}$ and $\varkappa=[(k_0k)/m\omega_0]E_0/E_{\text{cr}}$. Here, $(k_0k)=\omega_0\omega-\bm{k}_0\cdot\bm{k}$, with $k_0^{\mu}=(\omega_0,\bm{k}_0)$ and $k^{\mu}=(\omega,\bm{k})$ being the four-momentum of the laser photons and of the incoming photon, respectively. It is worth observing that in the so-called ``ultra-relativistic'' limit $\xi\to \infty$, the net number of laser photons absorbed in NBWPP is very large and of the order of $\xi^3$ \cite{Ritus_1985}. Since presently available optical lasers allow for values of $\xi$ of the order of $10^2$ \cite{Yanovsky_2008}, unprecedented degrees of nonlinearity of the order of one million are in principle achievable.

Refractive QED processes in a strong laser field involve only initial and final photons, and the background field \cite{Dittrich_b_2000}. Such processes of genuinely quantum nature are a unique tool for testing the predictions of strong-field QED on the nonlinear evolution of the electromagnetic field in vacuum. Vacuum polarization \cite{Baier_1976_b} and photon splitting \cite{Di_Piazza_2007} in a laser field are two examples of refractive QED processes, which have been considered in the literature. It has been observed in both cases, that the net number of laser photons exchanged with the laser field is very small (of the order of unity) even in the ultra-relativistic limit $\xi\to \infty$. As a general remark to be kept in mind throughout in the paper, we observe that the laser field is treated as a classical field in those papers and here as well. Thus, an expression like ``the net number of laser photons exchanged with the laser field is very small'' has to be intended more precisely as ``the net energy and momentum exchanged with the laser field is very small, corresponding to a few laser photons.''

In the present paper, by analyzing the amplitude of a general refractive QED effect, we indicate analytically that this is a general feature of such processes in a strong laser field. The physical origin of this effect lies in the fact that in a refractive QED process, the laser field couples to the external photons only indirectly via a virtual electron-positron pair. As we will see below, at higher and higher laser intensities the distance covered by the virtual electron and positron before annihilating decreases accordingly, in such a way that the process occurs with a net exchange of a low number of laser photons. This is in contrast, as we have mentioned, to the NBWPP, which is also primed in the collision of a (real) photon and a laser field. However, in NBWPP the final electron and positron are on the mass shell, requiring a large amount of laser photons to be absorbed for the process to occur at all in the presence of an ultra-relativistic laser field. Although the analysis is limited to the one-loop amplitude of a refractive QED effects and does not cover observable quantities as cross-sections or rates, the present results can be of relevance for future experimental campaigns, aiming to measure strong-field QED effects in the presence of a background laser field. As we will see, they indicate, for example, that, in order to detect refractive QED effects in a regime where higher-order effects in the laser-field amplitude are important, it is more convenient to measure the yield of final photons, rather than to measure the angular distribution or the energies of the final photons.

\section{Calculation of the amplitude of a generic refractive QED process}
Refractive QED processes in a laser field involve in general $N_i$ incoming, $N_o$ outgoing photons, with $N_i+N_o>1$, and the laser photons (the special case $N_i+N_o=1$, corresponding to the tadpole diagram, is trivial in the case of a background plane-wave field \cite{Schwinger_1951} and it will not be considered here). However, for the sake of notational simplicity, we consider here the abstract case of only incoming photons ($N_o=0$) and we set $N_i=N$. The external photons have momenta $k_j^{\mu}$ and polarization four-vectors $e_j^{\mu}$, with $j=1,\ldots,N$ (see Fig. 1): the $j$th incoming photon can be ``transformed'' into an outgoing one via the substitutions $k_j^{\mu}\to -k_j^{\mu}$ and $e_j^{\mu}\to e_j^{\mu\,*}$ in the amplitude (see Eq. (\ref{M}) below). As it will be clear below, the results of the paper are unaffected by this particular choice. Moreover, we limit here to the case of external real photons ($k_j^2=0$), although the analysis and the conclusions can be correspondingly extended to the case of off-shell external photons, as those representing external fields as, for example, a Coulomb field.
\begin{figure}
\begin{center}
\includegraphics[width=0.9\linewidth]{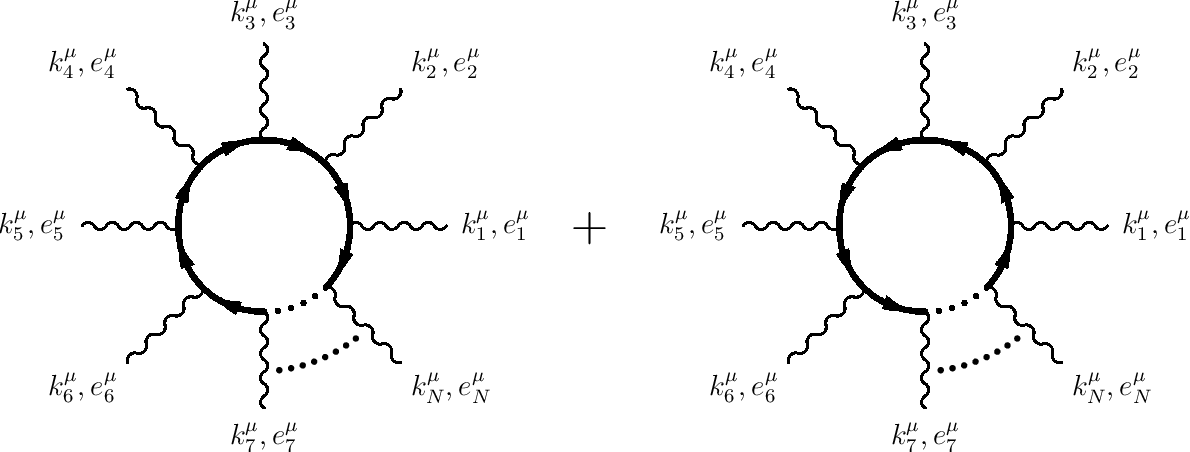}
\caption{Two typical Feynman diagrams relative to a generic refractive QED process in a laser field. The thin wavy lines indicate the external photons with four-momenta $k_1^{\mu},\ldots,k_N^{\mu}$ and polarization four-vectors $e_1^{\mu},\ldots,e_N^{\mu}$, respectively, and the thick plain lines indicate the laser-dressed electron propagators.}
\end{center}
\end{figure}
The mentioned process is described by the sum of all Feynman diagrams, which can be obtained from the one in the left side of Fig. 1 by permuting the labels in the photon legs. Among them, we consider here only the one in the right part of Fig. 1, and the treatment of the remaining diagrams can be performed in an analogous way (any diagram contributing to a refractive QED process can always be considered together with the other one, differing only in the direction of circulation of the four-momentum through the electron loop)\footnote{In the case $N=2$ the two diagrams in Fig. 1 coincide. Thus, if the amplitude $M$ is employed to calculate a rate, it has to be first divided by two to avoid over-counting.}. The reason for considering these two diagrams together is that this allows to formulate a simple set of substitution rules, which in turn clearly show the general structure of the amplitude of the process (see the discussions below Eq. (\ref{M_s}) here below and between Eqs. (\ref{T_p_3}) and (\ref{T_3_T}) in Appendix A). The amplitude $M$ corresponding to the diagrams in Fig. 1 is given by \cite{Landau_b_4_1982}
\begin{equation}
\label{M}
\begin{split}
M=&-e^N\int d^4x_1\cdots d^4x_Ne^{-i[(k_1x_1)+\cdots+(k_Nx_N)]}\\
&\times\text{Tr}[\hat{e}_1G(x_1,x_2|A)\hat{e}_2G(x_2,x_3|A)\cdots\hat{e}_NG(x_N,x_1|A)]+\circlearrowleft,
\end{split}
\end{equation}
where the ``hat'' indicates the contraction of a four-vector with the Dirac gamma matrices $\gamma^{\mu}$ and where the symbol $\circlearrowleft$ indicates the amplitude corresponding to the diagram on the right in Fig. 1. In Eq. (\ref{M}) the quantity $G(x,y|A)$ is the dressed electron propagator in the laser field. The latter is described by the four-vector potential $A^{\mu}=A^{\mu}(\phi)$, where $\phi=(nx)$, with $n^{\mu}=(1,\bm{n})$ and $\bm{n}$ being the propagation direction of the laser field. By working in the Lorentz gauge, the four-vector potential $A^{\mu}(\phi)$ of the laser field can be chosen in the form $A^{\mu}(\phi)=(0,\bm{A}(\phi))$, with $\bm{n}\cdot\bm{A}(\phi)=0$. Let $\bm{a}_1$ and $\bm{a}_2$ indicate the two possible independent laser polarization directions, such that $\bm{a}_r\cdot\bm{a}_s=\delta_{rs}$, with $r,s=1,2$, and that $\bm{a}_1\times\bm{a}_2=\bm{n}$. Then, the four-vector potential $A^{\mu}(\phi)$ can be written as  $A^{\mu}(\phi)=A_0[a^{\mu}_1\psi_1(\phi)+a^{\mu}_2\psi_2(\phi)]$, where $A_0=-E_0/\omega_0$, $a^{\mu}_r=(0,\bm{a}_r)$, and the two shape-functions $\psi_r(\phi)$ are arbitrary, smooth functions except that they satisfy the relation $\sqrt{\psi^{\prime 2}_1(\phi)+\psi^{\prime 2}_2(\phi)}\le 1$ for all values of $\phi$, with $\psi'_{1/2}(\phi)=d\psi_{1/2}(\phi)/d\phi$. Here, $E_0$ and $\omega_0$ indicate the laser-electric-field amplitude and its central angular frequency, respectively\footnote{More abstractly, but more in general, the quantity $\omega_0$ can be defined as a parameter characterizing the time dependence of the laser field and such that $\omega_0\phi$ is a dimensionless Lorentz scalar.}. Since the interaction of the $j$th photon with the laser field is controlled by the parameter $\varkappa_j=\eta_j\xi$, with $\eta_j=\omega_0k_{j,X}/m^2$ \cite{Ritus_1985,Di_Piazza_2012}, it is natural to assume here that $\varkappa_j\ne 0$ for all $j$s, which means $k_{j,X}\ne 0$ for all $j$s. This means that none of the external photons propagate along the same direction of the laser photons (of course, we exclude the trivial case of an external photon with vanishing energy).

In order to calculate the amplitude $M$, we employ below the operator technique, developed in \cite{Baier_1976_a,Baier_1976_b} for the case of a background plane-wave laser field (the calculation of the amplitude can of course also be performed by employing the standard Feynman rules in the Furry picture \cite{Landau_b_4_1982}, the advantage of the operator technique being to provide a more suitable expression of the amplitude to estimate the net number of photons exchanged with the laser field (see, in particular, the discussion below Eq. (\ref{Exp_2}))). In the operator technique the electron propagator in the laser field is written as $G(x,y|A)=\langle x|G(A)|y\rangle$, where
\begin{equation}
G(A)=\frac{1}{\hat{\Pi}-m+i\epsilon},
\end{equation}
with $\Pi^{\mu}=\Pi^{\mu}(A)=P^{\mu}-eA^{\mu}(\phi)$ and with $\epsilon$ being a positive infinitesimal quantity. Here, the four-vector $P^{\mu}$ is the four-momentum operator, satisfying the commutation rules $[x^{\mu},P^{\nu}]=-ig^{\mu\nu}$, where $g^{\mu\nu}=\text{diag}(+1,-1,-1,-1)$. By employing the above representation of the electron propagator and by using the cyclic property of the trace, the amplitude in Eq. (\ref{M}) can be simply written as
\begin{equation}
\label{M_op}
M=-e^N\int d^4x\,\text{Tr}\,\langle x\vert G_1(A)\cdots G_N(A)\vert x\rangle+\circlearrowleft,
\end{equation}
where we have introduced the block operators $G_j(A)=G(A)\hat{e}_j\exp[-i(k_jx)]$. It is convenient to express the amplitude $M$ in terms of the ``square'' propagator 
\begin{equation}
\label{D_0}
D(A)=\frac{1}{\hat{\Pi}^2-m^2+i\epsilon}
\end{equation}
rather than in terms of $G(A)$. The details of the procedure to carry this out are reported in the Appendix A. Here, we only provide a summary of this procedure in terms of substitution rules. The amplitude $M$, in fact, turns out to be expressed as
\begin{equation}
\label{M_s}
M=\frac{1}{2}\sum_{i=1}^{[N/2]+1}(M^{(i)}+\{1\ldots N\to N\ldots 1\}),
\end{equation}
where $M^{(i)}$ are partial amplitudes, with $[N/2]$ indicating the integer part of $N/2$. The quantity 
$\{1\ldots N\to N\ldots 1\}$ refers to the fact that each partial amplitude $M^{(i)}$ will have $N$ indexes corresponding to the $N$ ordered operators $G_j(A)$ in Eq. (\ref{M_op}), and it indicates that the same partial amplitude $M^{(i)}$ has to be added, but with the indexes $1,\ldots,N$ appearing in the opposite order $N,\ldots,1$. In turn, each partial amplitude $M^{(i)}$ is expressed as a sum $\sum_{J=1}^{J_i} M_J^{(i)}$ of terms $M_J^{(i)}$ and the number $J_i$ of terms in each partial amplitude depends on the partial amplitude itself. Each term $M_J^{(i)}$ has the form $-e^N\int d^4x\,\text{Tr}\,\langle x\vert O_J^{(i)}\vert x\rangle$, with the operator $O_J^{(i)}$ being obtained from the original operator product $G_1(A)\cdots G_N(A)$ by means of the following substitution rules: 
\begin{enumerate}
\item Partial amplitude $M^{(1)}$: substitute each block $G_j(A)$ by $D_j(A)\equiv D(A)\exp[-i(k_jx)][2(\Pi e_j)+\hat{k}_j\hat{e}_j]$ (this partial amplitude contains one term).
\item Partial amplitude $M^{(2)}$: combine two successive blocks $G_j(A) G_{j+1}(A)$ (for $j=1,\ldots, N$) and substitute this quantity with the ``contraction''\\ $-C_{j,j+1}(A)=-D(A)\hat{e}_j\exp[-i(k_jx)] \hat{e}_{j+1}\exp[-i(k_{j+1}x)]$, then substitute the remaining blocks as in 1.; it is understood that $G_{N+1}(A)\equiv G_1(A)$ and that $C_{N,N+1}(A)\equiv C_{N,1}(A)$; this partial amplitude contains $N$ terms.
\item Partial amplitude $M^{(3)}$: combine twice two successive blocks $G_j(A)G_{j+1}(A)$ and $G_{j'}(A)G_{j'+1}(A)$ (for $j=1,\ldots,N-2$, and for $j'=3,\ldots,N-1$ (if $j=1$) or for $j'=j+2,\ldots,N$ (if $j>1$)), and substitute these quantities with the contractions $-C_{j,j+1}(A)$ and $-C_{j',j'+1}(A)$, respectively; then substitute the remaining blocks as in 1.; it is understood that $G_{N+1}(A)\equiv G_1(A)$ and that $C_{N,N+1}(A)\equiv C_{N,1}(A)$; this partial amplitude has to be considered only if $N\ge 4$ and it contains $N(N-3)/2$ terms.
\item The above procedure continues by increasing by one the number of combinations of successive blocks. The last partial amplitude $M^{([N/2]+1)}$ contains the two terms $(-1)^{N/2}C_{1,2}(A)C_{3,4}(A)\cdots C_{N-1,N}(A)$ and \\$(-1)^{N/2}C_{N,1}(A)C_{2,3}(A)\cdots C_{N-2,N-1}(A)$ with $N/2$ contractions if $N$ is even, or the $N$ terms $(-1)^{(N-1)/2}D_1(A)C_{2,3}(A)\cdots C_{N-1,N}(A)$, \\ $(-1)^{(N-1)/2}C_{N,1}(A)D_2(A)C_{3,4}(A)\cdots C_{N-2,N-1}(A)$,..., \\$(-1)^{(N-1)/2}C_{1,2}(A)C_{3,4}(A)\cdots C_{N-2,N-1}(A)D_N(A)$ with $(N-1)/2$ contractions if $N$ is odd.
\end{enumerate}

Now, a useful exponential representation of the square propagator $D(A)$ has been found in \cite{Baier_1976_a,Baier_1976_b} (see also \cite{Di_Piazza_2007}\footnote{Due to a typographical misprint, the quantity $s$ is missing in the last exponent in Eq. (16) in \cite{Di_Piazza_2007}.}):
\begin{equation}
\label{D}
\begin{split}
D(A)=&-i\int_0^{\infty}ds\, e^{is(\hat{\Pi}^2-m^2+i\epsilon)}=-i\int_0^{\infty}ds\, e^{-i(m^2-i\epsilon)s}\\
&\times \bigg\{1+\frac{e\hat{n}}{2P_X}
[\hat{A}(\phi+2sP_X)-\hat{A}(\phi)]\bigg\}\\
&\times e^{-i\int_0^sds'\left[{\bm P}_\perp-e{\bm
A}(\phi+2s'P_X)\right]^2}e^{-2isP_\phi P_X},
\end{split}
\end{equation}
where we have introduced the operators $P_{\phi}=(P_t+P_{x_{\parallel}})/2$ and $P_X=-(P_t-P_{x_{\parallel}})=-(nP)$ of the conjugated momenta to the coordinates $\phi=t-x_{\parallel}$ and $X=(t+x_{\parallel})/2$, with $x_{\parallel}=\bm{n}\cdot\bm{x}$, such that $\phi$ and $X$ can be interpreted as a ``time'' and a ``space'' coordinate, respectively, i.e., $[\phi,P_{\phi}]=-i$ and $[X,P_{X}]=i$. Note that $t=X+\phi/2$, $x_{\parallel}=X-\phi/2$, $P_t=P_{\phi}-P_X/2$, and $P_{x_{\parallel}}=P_{\phi}+P_X/2$.

Out of the different partial amplitudes which arise from the above substitutions, we work out only the following one 
\begin{equation}
\label{M_op_1}
\begin{split}
M^{(1)}=&-e^N\int d^4x\,\text{Tr}\,\langle x|D(A)\text{e}^{-i(k_1x)}[2(\Pi e_1)+\hat{k}_1\hat{e}_1]\cdots \\
&\times D(A)\text{e}^{-i(k_Nx)}[2(\Pi e_N)+\hat{k}_N\hat{e}_N]|x\rangle,
\end{split}
\end{equation}
which arises from the substitution in 1.. This partial amplitude is always present, independently of the number of the external photons and, as it will also be clear from the considerations below, the analysis of the other partial amplitudes proceeds analogously. By looking at the expression of the operators $D(A)$ (see Eq. (\ref{D})), the coordinate operators $X$ and $\bm{x}_{\perp}$ appear to occur only in the exponentials relative to the external photons. By employing the operator identity $e^{i(k_jx)}f(P)e^{-i(k_jx)}=f(P+k_j)$, we can move all the operators $e^{i(k_{j,X}X+\bm{k}_{j,\perp}\cdot\bm{x}_{\perp})}$ to the left and let them act on the bra $\langle x|$. The result is
\begin{equation}
\label{M_op_2}
\begin{split}
M^{(1)}=&-e^N\int d^4x\,e^{i(K_XX+\bm{K}_{\perp}\cdot\bm{x}_{\perp})}\text{Tr}\,\langle x|e^{-ik_{1,\phi}\phi}\{2[(\Pi^{\mu}+\kappa^{\mu}_2) e_{1,\mu}]+\hat{k}_1\hat{e}_1\}D_2(A)\\
&\cdots\times e^{-ik_{N-1,\phi}\phi}\{2[(\Pi^{\mu}+\kappa^{\mu}_N) e_{N-1,\mu}]+\hat{k}_{N-1}\hat{e}_{N-1}\}D_N(A)\\
&\times e^{-ik_{N,\phi}\phi}[2(\Pi e_N)+\hat{k}_N\hat{e}_N]D(A)|x\rangle,
\end{split}
\end{equation}
where $K^{\mu}=\sum_{j=1}^Nk_j^{\mu}$, $\kappa_j^{\mu}=\sum_{i=j}^Nk_i^{\mu}$ (note that $\kappa_1^{\mu}=K^{\mu}$), and $D_l(A)=D(A)\vert_{P_X\to P_X+\kappa_{l,X},\bm{P}_{\perp}\to \bm{P}_{\perp}+\bm{\kappa}_{l,\perp}}$, with $l=2,\ldots,N$. Now, the operators between the bra $\langle x|$ and the ket $|x\rangle$ do not contain the coordinates $X$ and $\bm{x}_{\perp}$, and the identities
\begin{align}
\label{x_f_x}
\langle X|f(P_X)|X\rangle=\int\frac{dp_X}{2\pi}f(p_X), && \langle \bm{x}_{\perp}|g(\bm{P}_{\perp})|\bm{x}_{\perp}\rangle=\int\frac{d^2p_{\perp}}{(2\pi)^2}g(\bm{p}_{\perp})
\end{align}
valid for arbitrary functions $f(P_X)$ and $g(\bm{P}_{\perp})$ can be applied (we assumed here that the eigenstates $|p\rangle$ of the four-momentum operator $P^{\mu}$, i.e., $P^{\mu}|p\rangle=p^{\mu}|p\rangle$, are such that $\langle x\vert p\rangle=e^{-i(px)}$ and $\langle p|p'\rangle=(2\pi)^4\delta^4(p-p')$). Moreover, the integrals in $X$ and $\bm{x}_{\perp}$ are easily taken and the partial amplitude $M^{(1)}$ becomes
\begin{equation}
\label{M_op_3}
\begin{split}
M^{(1)}=&-(-ie)^N\delta(K_X)\delta^2(\bm{K}_{\perp})\int d\phi \int d p_X \int d^2 p_{\perp} \int_0^{\infty} ds_1\cdots ds_N\, e^{-i(m^2-i\epsilon)S} \\
&\times\text{Tr}\,\langle \phi|\{2[(p^{\mu}-eA^{\mu}(\phi)) e_{N,\mu}]+\hat{k}_N\hat{e}_N\}\\
&\times\left\{1+\frac{e}{2p_X}\hat{n}[\hat{A}(\phi+2s_1p_X))-\hat{A}(\phi)]\right\}\\
&\times e^{-i\int_0^{s_1}ds'_1[\bm{p}_{\perp}-e\bm{A}(\phi+2s'_1p_X)]^2}e^{-2is_1P_{\phi}p_X}e^{-i\kappa_{1,\phi}\phi}\\
&\times \{2[(p^{\mu}-eA^{\mu}(\phi)+\kappa^{\mu}_1) e_{1,\mu}]+\hat{k}_1\hat{e}_1\}\\
&\times\left\{1+\frac{e}{2(p_X+\kappa_{2,X})}\hat{n}[\hat{A}(\phi+2s_2(p_X+\kappa_{2,X}))-\hat{A}(\phi)]\right\}\\
&\times e^{-i\int_0^{s_2}ds'_2[\bm{p}_{\perp}+\bm{\kappa}_{2,\perp}-e\bm{A}(\phi+2s'_2(p_X+\kappa_{2,X}))]^2}e^{-2is_2P_{\phi}(p_X+\kappa_{2,X})}e^{-i\kappa_{2,\phi}\phi}\\
&\cdots\times \{2[(p^{\mu}-eA^{\mu}(\phi)+\kappa^{\mu}_{N-1}) e_{N-1,\mu}]+\hat{k}_{N-1}\hat{e}_{N-1}\}\\
&\times\left\{1+\frac{e}{2(p_X+\kappa_{N,X})}\hat{n}[\hat{A}(\phi+2s_N(p_X+\kappa_{N,X}))-\hat{A}(\phi)]\right\}\\
&\times e^{-i\int_0^{s_N}ds'_N[\bm{p}_{\perp}+\bm{\kappa}_{N,\perp}-e\bm{A}(\phi+2s'_N(p_X+\kappa_{N,X}))]^2}e^{-2is_NP_{\phi}(p_X+\kappa_{N,X})}e^{-i\kappa_{N,\phi}\phi}|\phi\rangle,
\end{split}
\end{equation}
where $S=s_1+\cdots+s_N$. We note that in this expression of the amplitude, we have substituted the operator $P^{\mu}$ with the number $p^{\mu}+\kappa^{\mu}_j$ in the four-dimensional scalar products $(Pe_j)$. First, we observe that, since $(k_je_j)=0$, then it is $(\kappa_je_j)=(\kappa_{j+1}e_j)$, for $j=1,\ldots,N-1$ and $(\kappa_Ne_N)=0$. Moreover, although the substitution $(Pe_j)\to (p^{\mu}+\kappa^{\mu}_j)e_{j,\mu}$ is evident for the components $p_X$ and $\bm{p}_{\perp}$ (see Eq. (\ref{x_f_x}) and the definition of the operators $D_l(A)$ below Eq. (\ref{M_op_2})), it is in principle not justified for the remaining component $P_{\phi}$. However, we show in the Appendix B that gauge invariance implies that the four-dimensional scalar products $(Pe_j)$ actually do not involve the component $P_{\phi}$. The remaining matrix element can be calculated by employing the identity
\begin{align}
e^{-i\phi_0P_\phi}|\phi\rangle=|\phi-\phi_0\rangle,
\end{align}
where $\phi_0$ is a constant, and the fact that $\langle \phi|\phi'\rangle=\delta(\phi-\phi')$. The resulting $\delta$-function 
$\delta(2s_1(p_X+\kappa_{1,X})+\cdots+2s_N(p_X+\kappa_{N,X}))$ can be exploited to perform the integral in $p_X$ and the result is
\begin{equation}
\label{M_op_4}
\begin{split}
M^{(1)}=&-\frac{(-ie)^N}{2}\delta(K_X)\delta^2(\bm{K}_{\perp})\int d\phi \int d^2 p_{\perp}\int_0^{\infty} \frac{ds_1\cdots ds_N}{S}\, e^{-i(m^2-i\epsilon)S} e^{-iK_{\phi}\phi}\\
&\times e^{-i\sum_{j=1}^N\int_0^{s_j}ds'_j\{\delta\kappa_{j,\phi}\delta\kappa_{j,X}+[\bm{p}_{\perp}+\bm{\pi}_{j,\perp}(\phi,s'_j)]^2\}}\\
&\times\text{Tr}\,\left\langle \{2[(p^{\mu}-eA^{\mu}(\phi)) e_{N,\mu}]+\hat{k}_N\hat{e}_N\}\left\{1+\frac{e}{2\delta\kappa_{1,X}}\hat{n}[\hat{A}(\phi+\phi_1)-\hat{A}(\phi)]\right\}\right.\\
&\times \{2[(p^{\mu}-eA^{\mu}(\phi+\phi_1)+\kappa^{\mu}_1) e_{1,\mu}]+\hat{k}_1\hat{e}_1\}\\
&\times\left\{1+\frac{e}{2\delta\kappa_{2,X}}\hat{n}[\hat{A}(\phi+\phi_2)-\hat{A}(\phi+\phi_1)]\right\}\\
&\cdots\times\{2[(p^{\mu}-eA^{\mu}(\phi+\phi_{N-1})+\kappa^{\mu}_{N-1}) e_{N-1,\mu}]+\hat{k}_{N-1}\hat{e}_{N-1}\}\\
&\left.\times\left\{1+\frac{e}{2\delta\kappa_{N,X}}\hat{n}[\hat{A}(\phi+\phi_N)-\hat{A}(\phi+\phi_{N-1})]\right\}\right\rangle.
\end{split}
\end{equation}
In this expression we have simplified the notation by introducing the ``average''
\begin{equation}
\label{Av}
\bar{f}=\frac{1}{S}\sum_{j=1}^N\int_0^{s_j}ds'_jf_j(s'_j)
\end{equation}
of $N$ arbitrary functions $f_j(s'_j)$, the residuals
\begin{equation}
\label{Res}
\delta f_j(s'_j)=f_j(s'_j)-\bar{f},
\end{equation}
and the quantities
\begin{equation}
\label{phi_l}
\phi_j=2\sum_{i=1}^j\delta\kappa_{i,X}s_i
\end{equation}
and
\begin{equation}
\label{pi_j}
\pi^{\mu}_j(\phi,s'_j)=\kappa_j^{\mu}-eA^{\mu}(\phi+\phi'_j),
\end{equation}
with
\begin{align}
\label{phi_p_1}
\phi'_1&=2\delta\kappa_{1,X}s'_1\\
\label{phi_p_l}
\phi'_l&=2\sum_{i=1}^{l-1}\delta\kappa_{i,X}s_i+2\delta\kappa_{l,X}s'_l, &&l=2,\ldots,N.
\end{align}
Note also that $p_X=-\bar{\kappa}_X$, that $\phi_N=0$ and that in our gauge $\pi_{j,X/\phi}(\phi,s'_j)=\kappa_{j,X/\phi}$. Moreover, in Eq. (\ref{M_op_4}) and in the successive expressions of $M^{(1)}$, the quantity $p_X$ in the trace has to be interpreted as $-\bar{\kappa}_X$.

In order to take the integral in $\bm{p}_{\perp}$, it is convenient first to shift $\bm{p}_{\perp}$ as $\bm{p}_{\perp}\to\bm{p}_{\perp}-\bar{\bm{\pi}}_{\perp}(\phi,\{s\})$, where $\{s\}=s_1,\ldots, s_N$. In this way, the resulting expression of the amplitude can be written as
\begin{equation}
\label{M_op_5}
\begin{split}
M^{(1)}=&-\frac{(-ie)^N}{2}\delta(K_X)\delta^2(\bm{K}_{\perp})\int d\phi \int d^2 p_{\perp}\int_0^{\infty} \frac{ds_1\cdots ds_N}{S}\, e^{-i[K_{\phi}\phi-F(\phi,\{s\})]}\\
&\times e^{-iS\bm{p}_{\perp}^2}\text{Tr}\bigg\langle\prod_{j=1}^N\{2[(p^{\mu}+\delta\pi_j^{\mu}(\phi,s_j)) e_{j,\mu}]+\hat{k}_j\hat{e}_j\}\\
&\left.\times\left\{1+\frac{e}{2\delta\kappa_{j+1,X}}\hat{n}[\hat{A}(\phi+\phi_{j+1})-\hat{A}(\phi+\phi_j)]\right\}\right\rangle,
\end{split}
\end{equation}
where
\begin{equation}
\label{F}
F(\phi,\{s\})=\sum_{j=1}^N\int_0^{s_j}ds'_j[\delta\pi^{\mu}_j(\phi,s'_j)\delta\pi_{j,\mu}(\phi,s'_j)-m^2+i\epsilon],
\end{equation}
where $\delta\kappa_{N+1}\equiv\delta\kappa_1$ and $\phi_{N+1}\equiv\phi_1$. The integral in $\bm{p}_{\perp}=(p_1,p_2)$ can be written as a sum of integrals of the form
\begin{equation}
I_{n_1,n_2}=\int d^2p_{\perp}\, p_1^{n_1}p_2^{n_2}\, e^{-iS\bm{p}_{\perp}^2},
\end{equation}
where $n_1$ and $n_2$ are two non-negative integers. The integral $I_{n_1,n_2}$ vanishes if $n_1$ and/or $n_2$ are odd, whereas it is equal to
\begin{equation}
I_{n_1,n_2}=2\pi\frac{(n_1-1)!!(n_2-1)!!}{(2iS)^{(n_1+n_2+2)/2}}
\end{equation}
if $n_1$ and $n_2$ are both even. In conclusion, we can write the partial amplitude $M^{(1)}$ in the compact form
\begin{equation}
\label{M_op_5_C}
\begin{split}
M^{(1)}=&\frac{i\pi}{2}(-ie)^N\delta(K_X)\delta^2(\bm{K}_{\perp})\int d\phi \int_0^{\infty} \frac{ds_1\cdots ds_N}{S^2}\, e^{-i[K_{\phi}\phi-F(\phi,\{s\})]}\\
&\times\text{Tr}\bigg\langle\prod_{j=1}^N\{2[(p^{\mu}+\delta\pi_j^{\mu}(\phi,s_j)) e_{j,\mu}]+\hat{k}_j\hat{e}_j\}\\
&\left.\times\left\{1+\frac{e}{2\delta\kappa_{j+1,X}}\hat{n}[\hat{A}(\phi+\phi_{j+1})-\hat{A}(\phi+\phi_j)]\right\}\right\rangle,
\end{split}
\end{equation}
where the substitution rules
\begin{align}
\label{Sub_X}
p_X\to&-\bar{\kappa}_X\\
\label{Sub_perp}
\left(\frac{(pa_1)}{\sqrt{-a_1^2}}\right)^{n_1}\left(\frac{(pa_2)}{\sqrt{-a_2^2}}\right)^{n_2}\to&
\begin{cases}
0 & \text{if $n_1$ and/or $n_2$ are odd}\\
\frac{(n_1-1)!!(n_2-1)!!}{(2iS)^{(n_1+n_2)/2}} & \text{if $n_1$ and $n_2$ are even}
\end{cases}
\end{align}
in the expression of the trace are understood. The amplitude in Eq. (\ref{M_op_5_C}) may diverge
for $N<5$ \cite{Landau_b_4_1982,Liang_2012}. The case $N=2$ (polarization operator) has been explicitly investigated
in \cite{Baier_1976_b} and the case $N=3$ has been considered in \cite{Di_Piazza_2007,Di_Piazza_2008_a}. The
regularization procedure can be carried out by first subtracting and adding the corresponding amplitude 
$M_0^{(1)}$ at zero external field, i.e. by writing $M^{(1)}=(M^{(1)}-M_0^{(1)})+M_0^{(1)}$. Gauge invariance ensures that the quantity $M^{(1)}-M_0^{(1)}$ is finite and that only the vacuum-term $M_0^{(1)}$ needs to be regularized (see, in particular, \cite{Baier_1976_b}). The same procedure can be applied to the remaining case $N=4$, where the divergences are in general
less severe than, e.g., for $N=2$. As it will be clear below, the present analysis is based essentially on the behavior of the field-dependent phase function $F(\phi,\{s\})$, then the conclusions, drawn starting from the unregularized amplitude $M^{(1)}$, also apply to the regularized one $M^{(1)}-M_0^{(1)}$. Since the regularization procedure is necessary only for $N<5$, in order to keep general the following formulas, we will still analyze the unregularized amplitude $M^{(1)}$, being understood, however, that for $N<5$, actually, the regularized amplitude $M^{(1)}-M_0^{(1)}$ has to be considered.

Before passing to the estimation of the net number of laser photon exchanged in a refractive QED process, we observe here that the integral representation
\begin{equation}
\label{Prop}
\prod_{j=1}^N\frac{1}{p_j^2-m^2+i\epsilon}=(-i)^N\int_0^{\infty}ds_1\cdots ds_N\,e^{i\sum_{j=1}^N\int_0^{s_j}ds'_j(p_j^2-m^2+i\epsilon)},
\end{equation}
of the electron propagator in vacuum in momentum space, suggests to interpret the quantity $\delta\pi^{\mu}_j(\phi,s'_j)$ as an ``effective'' instantaneous four-momentum of the virtual particle flowing between the $(j-1)$th and the $j$th vertex (see Eqs. (\ref{M_op_5_C}) and (\ref{F})).
%
%
\section{Estimation of the net number of exchanged laser photons}
\label{Estimation}
If there were no external laser field, the remaining integral in $\phi$ in Eq. (\ref{M_op_5_C}) would provide the $\delta$-function $\delta(K_{\phi})$, which, together with the other three $\delta$-functions, would imply the overall energy-momentum conservation $K^{\mu}=0$, as expected. In the presence of the laser field, a measure of the net number of photons exchanged with the laser field during the refractive QED process is determined by the quantity $K_{\phi}/\omega_0$, where $\omega_0$ is the central laser angular frequency. In order to estimate the net number of laser photons exchanged, we recall that the multiphoton nature of the process is controlled by the parameter $\xi=|e|E_0/m\omega_0$, where $E_0$ is the amplitude of the electric field of the laser \cite{Ritus_1985,Di_Piazza_2012}. From the physical meaning of this parameter, in fact, it is not surprising that if $\xi\lesssim 1$, the net number of photons exchanged with the laser field is of the order of unity. This regime is the relevant one for present and future x-ray laser facilities \cite{Di_Piazza_2012}, for which the parameter $\xi$ is not expected to exceed unity due to the relatively large photon energy ($\omega_0\gtrsim 1\;\text{KeV}$). Thus, we directly consider below the ultra-relativistic limit where $\xi\to \infty$, having in mind an optical laser system with $\omega_0\sim 1\;\text{eV}$. In order to further specify the physical regime, we have also to consider the parameters $\varkappa_j$ (see the discussion below Eq. (\ref{M})). If $\varkappa_j$ largely exceeds unity, an electron-positron pair can be in principle created in the collision of the laser field and the $j$th external photon. The subsequent emission of radiation by such a pair would represent a background for the refractive QED process. Thus, we limit here to the case where the parameters $\varkappa_j$ are fixed and less or of the order of unity, such that electron-positron pair production from laser-external photons is negligible. Correspondingly, we also exclude the possibility that electron-positron pairs can be created only by the external photons, even though, as it will be clear below, the following considerations will not depend formally on this condition.

It is convenient to write explicitly
\begin{equation}
\delta\pi^{\mu}_j(\phi,s'_j)\delta\pi_{j,\mu}(\phi,s'_j)=-2\delta\kappa_{j,X}\delta\kappa_{j,\phi}-[\delta\bm{\pi}_{j,\perp}(\phi,s'_j)]^2
\end{equation}
and to shift the variable $\phi$ as $\phi\to\phi+\Phi$, with $\Phi$ such that
\begin{equation}
\label{Shift}
K_{\phi}\Phi+2\sum_{j=1}^N\delta\kappa_{j,X}\delta\kappa_{j,\phi}s_j=0.
\end{equation}
In this way, the the partial amplitude $M^{(1)}$ can be written in the convenient form
\begin{equation}
\label{M_f}
\begin{split}
M^{(1)}=&\frac{i\pi}{2}(-ie)^N\delta(K_X)\delta^2(\bm{K}_{\perp})\int d\phi \int_0^{\infty} \frac{ds_1\cdots ds_N}{S^2}\, e^{-i[K_{\phi}\phi+F_{\phi}(\phi+\Phi,\{s\})]}\\
&\times\text{Tr}\bigg\langle\prod_{j=1}^N\{2[(p^{\mu}+\delta\pi_j^{\mu}(\phi+\Phi,s_j)) e_{j,\mu}]+\hat{k}_j\hat{e}_j\}\\
&\left.\times\left\{1+\frac{e}{2\delta\kappa_{j+1,X}}\hat{n}[\hat{A}(\phi+\Phi+\phi_{j+1})-\hat{A}(\phi+\Phi+\phi_j)]\right\}\right\rangle,
\end{split}
\end{equation}
where
\begin{equation}
\label{F_perp}
F_{\phi}(\phi+\Phi,\{s\})=\sum_{j=1}^N\int_0^{s_j}ds'_j\{[\delta\bm{\pi}_{j,\perp}(\phi+\Phi,s'_j)]^2+m^2-i\epsilon\}.
\end{equation}
The advantage of this form with respect to that in Eq. (\ref{M_op_5_C}) is that all the $N$ integrands in $F_{\phi}(\phi+\Phi,\{s\})$ are strictly positive and therefore that $F_{\phi}(\phi+\Phi,\{s\})\ge 0$. This implies, in fact, that the integration region in $ds_1\cdots ds_N$ mainly contributing to the partial amplitude $M^{(1)}$ is confined to sufficiently small values of $s_j$ such that that $F_{\phi}(\phi+\Phi,\{s\})\lesssim 1$, as otherwise the function $\exp(-iF_{\phi}(\phi+\Phi,\{s\}))$ would be highly oscillating. From what we mentioned at the beginning of this section, this would already indicate that the net number of photon exchanged during the refractive QED process is of the order of unity. However, in order to complete the proof, we have still to analyze the pre-exponential function. In fact, if $N$ is small, then the different powers of the external field present in this function would not essentially change the net number of laser photons exchanged. However, this could in principle occur for large $N$s. In order to show that this is not the case, we recall that in the considered regime, the parameters $\eta_j=\varkappa_j/\xi$ are much smaller than unity and therefore, in the effective integration region with respect to the variables $s_1,\ldots,s_N$, it is $\omega_0|\delta\kappa_{j,X}|s_j\lesssim\omega_0|\delta\kappa_{j,X}|/m^2\ll 1$, where we used the fact that $s_j\lesssim 1/m^2$ (see Eq. (\ref{F_perp})). Consequently, it results that $\omega_0|\phi_j|,\omega_0|\phi'_j|\ll 1$ and, by assuming that  $|k_{j,\phi}|\lesssim |K_{\phi}|$ for all $j$s, that $\omega_0|\Phi|\ll 1$ (see Eq. (\ref{Shift})). This observation allows one to expand the four-vector potential in Eq. (\ref{M_f}) as\footnote{We note that the above expansions also hold for larger values of the parameters $\varkappa_j$. In fact, instead of assuming that the parameters $\eta_j$ are much smaller than unity as in the text, we assume here that they are such that $\delta \kappa_{j,X}s_j\sim 1$. In this case, one cannot perform the mentioned expansions and the condition $F_{\phi}(\phi+\Phi,\{s\})\lesssim 1$ would imply that $s_j\lesssim 1/m^2\xi^2$. In turn, the condition $\delta \kappa_{j,X}s_j\sim 1$ would require that $\varkappa_j\sim \xi^3$. However, since it is assumed that $\xi\gg 1$, then in order the mentioned expansions not to be valid, it should be $\varkappa_j\sim 10^3$, where even the perturbative approach in the photon-electron interaction in QED in the presence of the laser field would break down \cite{Ritus_1985} (see also the discussion at the end of sec. \ref{Discussion}).}
\begin{align}
A^{\mu}(\phi+\Phi+\phi_j)\approx &A^{\mu}(\phi)-2E^{\mu}(\phi)\bigg(\Phi+\sum_{i=1}^j\delta\kappa_{i,X}s_i\bigg)\\
A^{\mu}(\phi+\Phi+\phi'_j)\approx &A^{\mu}(\phi)-2E^{\mu}(\phi)\bigg(\Phi+\sum_{i=1}^{j-1}\delta\kappa_{i,X}s_i+\delta\kappa_{j,X}s'_j\bigg),
\end{align}
where $E^{\mu}(\phi)=-dA^{\mu}(\phi)/d\phi$ (note that $E^{\mu}(\phi)$ is not a four-vector). Analogously, one obtains
\begin{align}
\label{Exp_1}
\hat{A}(\phi+\Phi+\phi_{j+1})-\hat{A}(\phi+\Phi+\phi_j)\approx &-2\hat{E}(\phi)\delta\kappa_{j+1,X}s_{j+1}\\
\label{Exp_2}
\begin{split}
\delta\pi_j^{\mu}(\phi+\Phi,s'_j)\approx &\delta \kappa_j^{\mu}+2eE^{\mu}(\phi)\bigg[\sum_{i=1}^{j-1}\delta\kappa_{i,X}s_i+\delta\kappa_{j,X}s'_j\\
&-\frac{1}{S}\sum_{l=1}^Ns_l\bigg(\sum_{i=1}^{l-1}\delta\kappa_{i,X}s_i+\frac{1}{2}\delta\kappa_{l,X}s_l\bigg)\bigg].
\end{split}
\end{align}
Now, the fact that $F_{\phi}(\phi+\Phi,\{s\})\lesssim 1$ implies, as an order-of-magnitude estimate, that $[\delta\bm{\pi}_{j,\perp}(\phi+\Phi,s_j)]^2s_j\lesssim 1/N$. Thus, the above expansions, together with the fact that $|\bm{p}_{\perp}|\sim 1/\sqrt{S}$ (see Eq. (\ref{Sub_perp})), indicate that in the effective formation region of the process, the ratio between the terms in the pre-exponent proportional to the laser field and those which do not contain the laser field itself is less than unity. Therefore, terms containing higher powers of the external field are subdominant and, in conclusion, the probability of an exchange of a net number of photons much larger than unity is suppressed also for large values of $N$.

In order to make our analysis more concrete, we consider the particular case of a monochromatic, circularly polarized laser field. In this case, the vector potential is given by $\bm{A}(\phi)=-(E_0/\omega_0)[\cos(\omega_0\phi)\bm{a}_1+\sin(\omega_0\phi)\bm{a}_2]$. Starting again from the general expression in Eq. (\ref{M_f}) (see also Eq. (\ref{F_perp})), it is convenient to introduce the vectors 
\begin{align}
\bm{a}_{j,c}(s'_j)&=C_j(s'_j)\bm{a}_1+S_j(s'_j)\bm{a}_2\\
\bm{a}_{j,s}(s'_j)&=-S_j(s'_j)\bm{a}_1+C_j(s'_j)\bm{a}_2,
\end{align}
where  $C_j(s'_j)=\cos(\omega_0(\Phi+\phi'_j))$ and $S_j(s'_j)=\sin(\omega_0(\Phi+\phi'_j))$. In this way, we obtain 
\begin{equation}
\delta\bm{\pi}_{j,\perp}(\phi+\Phi,s'_j)=\delta\bm{\kappa}_{j,\perp}-m\xi[\cos(\omega_0\phi)\delta\bm{a}_{j,c}(s'_j)+\sin(\omega_0\phi)\delta\bm{a}_{j,s}(s'_j)]
\end{equation}
and the function $F_{\phi}(\phi+\Phi,\{s\})$ can be written as 
\begin{equation}
F_{\phi}(\phi+\Phi,\{s\})=F_0(\{s\})+F_c(\{s\})\cos(\omega_0\phi)+F_s(\{s\})\sin(\omega_0\phi),
\end{equation}
where
\begin{align}
F_0(\{s\})=&\sum_{j=1}^N\int_0^{s_j}ds'_j\bm{(}(\delta\bm{\kappa}_{j,\perp})^2+m^2\{1+\xi^2[(\delta C_j(s'_j))^2+(\delta S_j(s'_j))^2]\}-i\epsilon\bm{)},\\
F_{c/s}(\{s\})=&-2m\xi\sum_{j=1}^N\int_0^{s_j}ds'_j\delta\bm{\kappa}_{j,\perp}\cdot\delta\bm{a}_{j,c/s}(s'_j).
\end{align}
Note that the integrals in $ds'_j$ in $F_0(\{s\})$ and $F_{c/s}(\{s\})$ can be easily taken in the present case, which is however not necessary here. The discussion below Eq. (\ref{F_perp}) indicates that in the effective integration region it is $F_0(\{s\}),|F_{c/s}(\{s\})|\lesssim 1$. We consider now the prototype integral in $\phi$
\begin{equation}
\label{I}
\mathcal{I}(\{s\})=\int d\phi\, \text{e}^{-i[K_{\phi}\phi+F_{\phi}(\phi+\Phi,\{s\})]},
\end{equation}
which is present in the partial amplitude $M^{(1)}$. After introducing the quantities $F_A(\{s\})$ and $\varphi_0(\{s\})$ according to the definitions 
\begin{align}
F_c(\{s\})&=F_A(\{s\})\cos(\varphi_0(\{s\})),\\
F_s(\{s\})&=F_A(\{s\})\sin(\varphi_0(\{s\})),
\end{align}
and after passing to the variable $\varphi=\omega_0\phi-\varphi_0(\{s\})$, we obtain
\begin{equation}
\label{I_f}
\mathcal{I}(\{s\})=2\pi \text{e}^{-i[(K_{\phi}/\omega_0)\varphi_0(\{s\})+F_0 (\{s\})]}\sum_{n_l=-\infty}^{\infty}i^{-n_l}\delta(K_{\phi}-n_l\omega_0)J_{n_l}(F_A(\{s\})),
\end{equation}
where we employed the identity $\exp(iz\cos\varphi)=\sum_{n=-\infty}^{\infty}i^nJ_n(z)\exp(in\varphi)$ in terms of the ordinary Bessel functions $J_n(z)$ of integer order $n$, valid for an arbitrary complex number $z$ \cite{Gradshteyn_b_2000}. Equation (\ref{I_f}) shows that $n_l$ indicates the net number of photons absorbed from (if $n_l<0$) or ceded to (if $n_l>0$) the laser field. The well-known property of ordinary Bessel functions $J_n(x)$ of a real (positive) argument of being much smaller than unity at $n\gg x$ and the fact that $F_A(\{s\})=\sqrt{F^2_c(\{s\})+F^2_s(\{s\})}\lesssim 1$ shows, at least for the terms in the pre-exponent independent of the laser field, that the net number of photons exchanged with the laser field is of the order of unity. The general observation below Eq. (\ref{Exp_2}) indicates that also high-order terms in the laser field in the pre-exponential will not essentially increase the net number of laser photons exchanged during the refractive QED process. Note that the fact that only a low net number of laser photons are exchanged during a refractive QED effects implies that the strong background laser field is practically not altered by the process itself. This is in agreement with the use here of the Furry picture, which includes the external field as a ``given'' field.

Before discussing the obtained results, it is worth observing that in the special case where $N=2$ and with two external real photons the net exchange of laser photons is exactly zero, due to \emph{kinematical} reasons \cite{Baier_1976_b,Dittrich_b_2000}. Our results show that there is a \emph{dynamical} reason such that the net exchange of laser photons is small also for arbitrary $N$.

%
%
\section{Discussion}
\label{Discussion}
As we have already mentioned above, it is interesting to compare the low net exchange of laser photon in a refractive QED process with what happens in the case of the NBWPP, which does also occur in the collision of a real photon and a laser field. Again, we limit in particular to the strong-field limit corresponding to $\xi\gg 1$ at fixed invariant parameters $\varkappa_j\sim 1$. The real electron and positron created via the NBWPP at $\xi\gg 1$ are already ultra-relativistic and a large net number of laser photons of the order of $\xi^3$ are absorbed from the laser field in order to fulfill energy-momentum conservation \cite{Ritus_1985}. On the other hand, a refractive QED process occurs via a virtual electron-positron pair and this manifests itself in the appearance of the integrals in $ds_1\cdots ds_N$ in the partial amplitude $M^{(1)}$. At larger and larger values of the electric field amplitude, the effective integration region in $ds_1\cdots ds_N$ reduces accordingly, in such a way that the function $F_{\phi}(\phi+\Phi,\{s\})$ is always of the order of or less than unity, and then that the net number of laser photons exchanged is of the order of unity, too. More specifically, we recall that if $p^{\mu}=(\varepsilon,\bm{p})$ is the momentum of a classical electron at the initial value $\phi=0$ ($\bm{A}(0)=\bm{0}$), then the component $p_{\phi}(\phi)$ of the four-momentum $p^{\mu}(\phi)=(\varepsilon(\phi),\bm{p}(\phi))$ at $\phi$ is given by \cite{Landau_b_2_1975}
\begin{equation}
p_{\phi}(\phi)=-\frac{m^2+[\bm{p}_{\perp}-e\bm{A}(\phi)]^2}{2p_X}.
\end{equation}
By performing the change of variable $\phi'_j=2\delta\kappa_{j,X}s'_j$ in Eq. (\ref{F_perp}), we see that $F_{\phi}(\phi+\Phi,\{s\})$ qualitatively corresponds to the quantity $\sum_{j=1}^N\int_{\phi_{j-1}}^{\phi_j}d\phi'_j\mathcal{P}_{j,\phi}(\phi'_j)$, where $\phi_0=0$ and where $\mathcal{P}_{j,\phi}(\phi'_j)$ is the component $\phi$ of the four-momentum of the virtual electron/positron flowing between the $(j-1)$th vertex and the $j$th vertex. Thus, the condition $F_{\phi}(\phi+\Phi,\{s\})\lesssim 1$ corresponds to the fact that, according to Heisenberg uncertainty principle, the virtual electron-positron pair annihilates after an interval $\Delta\phi'_j$ in $\phi'_j$ given by $\Delta\phi'_j\sim 1/\mathcal{P}_{\phi,j}$, where $\mathcal{P}_{\phi,j}$ indicates the order of magnitude of the momentum flowing between the $(j-1)$th vertex and the $j$th vertex. This corroborates the interpretation that in a refractive QED process, the stronger is the laser field, the higher is the four-momentum flowing through the electron-positron loop. Accordingly, the virtual electron-positron pair propagates for a shorter distance inside the laser field, such that the net number of photons, that can be exchanged in the process is always of the order of unity.

This difference between the net number of photons exchanged with the laser field in a general refractive QED process, inferred here from the investigation of the amplitude of such processes, and in NBWPP could appear at first sight not to be compatible with the optical theorem, when the imaginary part of the (reduced) amplitude of a refractive QED process can be related to the total rate of the corresponding pair-production process (e.g., the refractive QED process corresponding to NBWPP is essentially the polarization operator) \cite{Landau_b_4_1982}. However, this is not the case, because the \emph{total} rate of a pair-production process does not contain information on the net number of photons exchanged with the laser field, as the rate is integrated over the whole phase space of the created electron and positron. More quantitatively, since a plane-wave field depends only on the spacetime variable $\phi$, it is possible to write th $S$-matrix element $S_{fi}$ of an arbitrary process occurring in such a background field as
\begin{equation}
S_{fi}=\delta_{fi}+i(2\pi)^3\delta^2(\bm{P}_{f,\perp}-\bm{P}_{i,\perp})\delta(P_{f,X}-P_{i,X})R_{fi},
\end{equation}
where $P_{i/f}^{\mu}$ indicates the total initial/final four-momentum. The optical theorem \cite{Landau_b_4_1982} here reads
\begin{equation}
2\,\text{Im}(R_{ii})=\sum_f(2\pi)^3\delta^2(\bm{P}_{f,\perp}-\bm{P}_{i,\perp})\delta(P_{f,X}-P_{i,X})|R_{fi}|^2
\end{equation}
and we are interested to the case in which in the initial state there are a certain number of photons, whereas in the final state an electron-positron pair is present. By limiting, for simplicity, to the case of a monochromatic laser field with angular frequency $\omega_0$, we can expand the amplitude $R_{fi}$ as
\begin{equation}
\label{R_fi}
R_{fi}=\sum_{n_l=-\infty}^{\infty}(2\pi)\delta(P_{f,\phi}-P_{i,\phi}-n_l\omega_0)T_{n_l,fi},
\end{equation}
and the optical theorem provides the relation
\begin{equation}
2\,\text{Im}(T_{0,ii})=\sum_{n_l=-\infty}^{\infty}\sum_f(2\pi)^4\delta(P_f^{\mu}-P_i^{\mu}-n_l\omega_0n^{\mu})|T_{n_l,fi}|^2.
\end{equation}
On the one hand, this identity shows that only the quantity $T_{0,ii}$ corresponding to no net exchange of laser photons in a refractive QED process is relevant for the optical theorem. On the other hand, as already mentioned, all the quantities $|T_{n_l,fi}|^2$ corresponding to a given net exchange of an arbitrary number of laser photons in the pair-production process are summed up in the right-hand side of Eq. (\ref{R_fi}), in such a way that the resulting quantity does not contain any information on the typical number of laser photons net-exchanged during the process. In the specific example of NBWPP, the above conclusion is confirmed by the fact that the total pair production rate at $\xi\gg 1$ becomes independent of the parameter $\xi$ (it depends only on the parameter $\varkappa=(\omega_0k_X/m^2)(E_0/E_{cr})$, where $k^{\mu}$ is the four-momentum of the external photon), and it coincides with the corresponding total rate in a ``phase-dependent'' constant-crossed field but averaged over the laser phase \cite{Ritus_1985}. 

It is also worth observing that, although, according to the analysis above of the amplitude of a refractive QED process, the net number of laser photons exchanged in such a process is of the order of unity, high-order terms in the laser field amplitude contribute to the process (as, for example, in the Bessel functions in Eq. (\ref{I_f})). Such nonlinear terms stem for the exchange of laser photons without a net absorption or emission during the process. The fact that $F_{\phi}(\phi+\Phi,\{s\})\lesssim 1$ (that $F_A(\{s\})\lesssim 1$ in Eq. (\ref{I_f}) for the case of a circularly-polarized, monochromatic laser field) suggests that in general the exchange of a large number of laser photons is suppressed. At the same time, however, such higher-order nonlinear effects can strongly modify the amplitude of a refractive QED process. This observation suggests that, in general, in order to detect higher-order nonlinear effects in the laser amplitude in a refractive QED process, it is more convenient to measure yields of final photons, rather than to measure, for example, the energy or the angular distribution of the final photons (note that refractive QED processes involving an odd number of external photons cannot occur in vacuum, i.e., in the absence of any background field, due to parity conservation (Furry theorem \cite{Furry_1937})). In fact, the optimal regime of parameters to detect higher-order nonlinear effects in the laser-field amplitude in a refractive QED process is at $\varkappa_j\sim 1$, as $if \varkappa_j\ll 1$ the amplitude is approximately equal to its corresponding expression including only the leading-order term(s) in $\varkappa_j$. Now, even considering next generation of 10-PW optical laser systems \cite{Di_Piazza_2012}, providing an intensity of the order of $10^{23}\;\text{W/cm$^2$}$, the ratio $E_0/E_{\text{cr}}$ is smaller that $5\times 10^{-4}$. Thus, in order to have $\varkappa_j\sim 1$, initial photon energies are required of the order of $1\;\text{GeV}$. For final photon energies of this order of magnitude, if only a few photons from an optical laser ($\omega_0\sim 1\;\text{eV}$) are effectively exchanged, it is not feasible in practice at $\varkappa_j\sim 1$ to detect higher-order effects in the laser-field amplitude by measuring the final photons' energies and/or angular distribution (note that the typical energy and angular resolutions of electromagnetic calorimeters in the GeV range are of the order of 100 MeV and of a few mrad, respectively \cite{CMS_ECAL}, whereas the energy and the angle resolutions required here would be of the order of $\omega_0\sim 1\;\text{eV}$ and of $\omega_0/1\;\text{GeV}\sim 10^{-9}\;\text{rad}$, respectively). On the other hand, at $\varkappa_j\sim 1$ the amplitude of a refractive QED effect is expected to be substantially altered by higher-order terms in $\varkappa_j$ (see, for example, the Bessel functions in Eq. (\ref{I_f})), indicating that the measurement of the photon yield could be a more convenient observable to detect such higher-order effects. However, since the above discussion does not contain an estimate of the expected cross section or rate of a general refractive QED process, it cannot be considered as an experimental proposal but rather as an observation on what it could be convenient to measure, in order to detect higher-order nonlinear effects in refractive QED effects. If one is not interested in detecting higher-order effects in the laser-field amplitude, one can also allow for $\varkappa_j\ll 1$ and try to measure only leading-order effects. In fact, there are already more concrete suggestions in order to detect leading-order refractive QED effects at $N=2$ (vacuum polarization effects), e.g., by measuring the change in polarization of a probe photon passing through a laser field \cite{Heinzl_2006,Di_Piazza_2006}, or by directly detecting photon-photon scattering \cite{Bernard_2000,Lundstroem_2006,Tommasini_2008,King_2010,Kryuchkyan_2011} (see \cite{Di_Piazza_2012} for a more complete review on such experimental suggestions). We also shortly mention analogous experiments to detect vacuum polarization effects in a magnetic field \cite{Bregant_2008,Zavattini_2012} and in waveguides \cite{Brodin_2001}. The mentioned experiments employ low-energy photons (optical and/or x-ray) such that they are not suitable to detect \emph{higher-order} nonlinear effects in the laser field, because, in the notation of the present manuscript, $\varkappa_j\ll 1$ there. However, this does not imply that the processes themselves cannot be observed. On the contrary, it has already been noticed (see, e.g., \cite{Varfolomeev_1966,Bernard_2000,Lundstroem_2006}) that, employing intense optical lasers, leads to a large enhancement of the photon-photon scattering signal, by exploiting the stimulated emission of a photon in the presence of a large number of photons in the same mode.

In the analysis carried out so far, it has been assumed that radiative corrections are negligible. In the presence of an ultra-relativistic external plane-wave field this is the case if $\alpha\varkappa_j^{2/3}\ll 1$ for all $j$, where $\alpha=e^2\approx 1/137$ is the fine-structure constant, i.e., if $\varkappa_j\ll 1/\alpha^{3/2}\approx 10^3$ \cite{Ritus_1985}. However, radiative corrections and high-order diagrams would in any case involve only virtual particles in such a way that the physical argument given above and concerning the net number of laser photon exchanged would again apply. On the other hand, as we have already mentioned, the regime $\varkappa_j\gg 1$ is not suitable for observing a refractive QED process, due to the background photons emitted by the produced electron-positron pairs.

\section{Conclusions}
In conclusion, by employing the operator technique, we have shown that refractive QED processes in a laser field are likely to occur with a net absorption/emission of only a few laser photons even in the ultra-relativistic regime $\xi\gg 1$. The above analysis has been carried out only on the one-loop amplitude of a general refractive QED process and, for a final, conclusive answer, observables as the cross sections or the rates should be investigated. However, the present investigation can be already of relevance for experimental campaigns at future laser facilities. On this respect, our main conclusion is that in order to experimentally observe higher-order nonlinear effects in the laser-field amplitude in such processes, it is more convenient to measure yields of final photons in a refractive QED process, rather than, for example, to measure the energies or the angular distribution of the final photons.

\section*{Acknowledgments}
The author is grateful to K. Z. Hatsagortsyan, S. Meuren, and A. I. Milstein for useful discussions and to C. H. Keitel and F. Mackenroth for reading the manuscript.

\appendix

\section{}
In the present appendix we will indicate how to express the amplitude (\ref{M_op}) in such a way that it contains only the square propagators $D(A)$ (see Eq. (\ref{D_0})). It is convenient to introduce here the notation (note that some of the above symbols have been already introduced between Eq. (\ref{M_op}) and Eq. (\ref{D}))
\begin{align}
G_j(A)=&G(A)\hat{e}_j\exp[-i(k_jx)],\\
D_j(A)=&D(A)\exp[-i(k_jx)][2(\Pi e_j)+\hat{k}_j\hat{e}_j],\\
Q_j(A)=&D(A)\hat{e}_j\exp[-i(k_jx)]G^{-1}(A),\\
C_{j,j+1}(A)=&D(A)\hat{e}_j\exp[-i(k_jx)] \hat{e}_{j+1}\exp[-i(k_{j+1}x)].
\end{align}
The following identities can be easily proven
\begin{align}
\label{GDQ}
G_j(A)=&D_j(A)-Q_j(A),\\
\label{QD}
Q_j(A)D_{j+1}(A)=&Q_j(A)Q_{j+1}(A)+C_{j,j+1}(A),
\end{align}
where for $j=N$, the index $N+1$ has to be intended as $1$ (recall the cyclic property of the trace). In order to further simplify the notation, we
also define the generalized trace of a matrix operator $O$
\begin{equation}
\text{Tr}_x(O)=\int d^4x \text{Tr}\langle x|O|x\rangle
\end{equation}
such that it is sufficient to analyze the quantity
\begin{equation}
T_N(A)=\text{Tr}_x[G_1(A)\cdots G_N(A)]+\circlearrowleft.
\end{equation}
Since, as will be clear, the procedure to transform the quantity $T_N(A)$ only depends on if $N$ is odd or even, we explicitly work out only the cases $N=3$ and $N=4$, being the cases $N>4$ completely analogous. Now,
\begin{equation}
\begin{split}
T_3(A)&=\text{Tr}_x[G_1(A)G_2(A)G_3(A)]+\circlearrowleft\\
&=\text{Tr}_x[(D_1(A)-Q_1(A))(D_2(A)-Q_2(A))(D_3(A)-Q_3(A))]+\circlearrowleft\\
&=\text{Tr}_x[D_1(A)D_2(A)D_3(A)]-\text{Tr}_x[Q_1(A)D_2(A)D_3(A)]\\
&\quad-\text{Tr}_x[D_1(A)Q_2(A)D_3(A)]-\text{Tr}_x[D_1(A)D_2(A)Q_3(A)]\\
&\quad+\text{Tr}_x[D_1(A)Q_2(A)Q_3(A)]+\text{Tr}_x[Q_1(A)D_2(A)Q_3(A)]\\
&\quad+\text{Tr}_x[Q_1(A)Q_2(A)D_3(A)]-\text{Tr}_x[Q_1(A)Q_2(A)Q_3(A)]+\circlearrowleft.\\
\end{split}
\end{equation}
The first term already contains only square propagators and, by applying the identity (\ref{QD}) to the three terms containing only one operator $Q_j(A)$, we see that the contributions coming from the first term in Eq. (\ref{QD}) exactly cancel the terms containing two operators $Q_j(A)Q_{j+1}(A)$. Thus, we obtain
\begin{equation}
\begin{split}
T_3(A)&=\text{Tr}_x[D_1(A)D_2(A)D_3(A)]-\text{Tr}_x[C_{1,2}(A)D_3(A)]-\text{Tr}_x[D_1(A)C_{2,3}(A)]\\
&\quad-\text{Tr}_x[C_{3,1}(A)D_2(A)]-\text{Tr}_x[Q_1(A)Q_2(A)Q_3(A)]+\circlearrowleft.
\end{split}
\end{equation}
Now, we consider separately the quantity
\begin{equation}
\label{T_p_3}
\begin{split}
T_{+,3}(A)&=\text{Tr}_x[Q_1(A)Q_2(A)Q_3(A)]\\
&=\text{Tr}_x\left[G_+(A)\hat{e}_1e^{-i(k_1x)}G_+(A)\hat{e}_2e^{-i(k_2x)}G_+(A)\hat{e}_3e^{-i(k_3x)}\right],
\end{split}
\end{equation}
where we have introduced the quantity $G_+(A)=(\hat{\Pi}+m+i\epsilon)^{-1}$, which corresponds to the electron propagator but with $m\to -m$. By imagining to work in the Dirac representation of the gamma matrices \cite{Landau_b_4_1982}, we consider the unitary matrix $U=\gamma^0\gamma^2\gamma^5$ and we note that $U\gamma^{\mu}U^{\dag}=\gamma^{\mu,t}$, where the upper index $t$ indicates the transpose with respect to the Dirac-matrices indexes. Since the four-momentum operator is hermitian, it is easy to show that $UG_+(A)U^{\dag}=-[G(-A)]^{t_x}$, where the upper index $t_x$ indicates the transpose with respect to the Dirac-matrices and to the spacetime indexes. In this way, by inserting the unity operator $UU^{\dag}$ in Eq. (\ref{T_p_3}) before and after each $\hat{e}_j$ and by exploiting the fact that $\text{Tr}_x(O^{t_x}_1O^{t_x}_2)=\text{Tr}_x[(O_2O_1)^{t_x}]=\text{Tr}_x(O_2O_1)$ for arbitrary operators $O_1$ and $O_2$, we obtain
\begin{equation}
\begin{split}
T_{+,3}(A)=&\text{Tr}_x[Q_1(A)Q_2(A)Q_3(A)]=\\
=&-\text{Tr}_x\left[G(-A)\hat{e}_3e^{-i(k_3x)}G(-A)\hat{e}_2e^{-i(k_2x)}G(-A)\hat{e}_1e^{-i(k_1x)}\right]\\
=&-\text{Tr}_x[G_3(-A)G_2(-A)G_1(-A)].
\end{split}
\end{equation}
Now, we recall that, in general, the quantity $T_N(A)$ also contain the contribution from the Feynman diagram where the electron arrows are reversed (see Fig. 1) and that, due to Furry theorem \cite{Landau_b_4_1982}, only terms proportional to an odd power of laser amplitude effectively contribute to $T_3(A)$, i.e., $T_3(A)=-T_3(-A)$. Therefore, by applying the same above procedure to the additional contribution from the Feynman diagram where the electron arrows are reversed, we obtain
\begin{equation}
\label{T_3_T}
\begin{split}
T_3(A)=&\frac{1}{2}\{\text{Tr}_x[D_1(A)D_2(A)D_3(A)]-\text{Tr}_x[C_{1,2}(A)D_3(A)]\\
&-\text{Tr}_x[D_1(A)C_{2,3}(A)]-\text{Tr}_x[C_{3,1}(A)D_2(A)]+\{123\to 321\})\},
\end{split}
\end{equation}
where the quantity $\{123\to 321\}$ means that the previous terms have to be added, but with the indexes $1,2$ and $3$ appearing in the opposite order $3,2$ and $1$. This result exactly corresponds to the general procedure given in the main text below Eq. (\ref{M_op}) for the case $N=3$.

The case with $N=4$ can be worked out in a completely analogous way and we only stress the differences with respect to the case $N=3$. The starting point is the quantity
\begin{equation}
\begin{split}
T_4(A)&=\text{Tr}_x[G_1(A)G_2(A)G_3(A)G_4(A)]+\circlearrowleft\\
&=\text{Tr}_x[(D_1(A)-Q_1(A))(D_2(A)-Q_2(A))(D_3(A)-Q_3(A))\\
&\quad\times(D_4(A)-Q_4(A))]+\circlearrowleft\\
&=\text{Tr}_x[D_1(A)D_2(A)D_3(A)D_4(A)]-\text{Tr}_x[Q_1(A)D_2(A)D_3(A)D_4(A)]\\
&\quad-\text{Tr}_x[D_1(A)Q_2(A)D_3(A)D_4(A)]-\text{Tr}_x[D_1(A)D_2(A)Q_3(A)D_4(A)]\\
&\quad-\text{Tr}_x[D_1(A)D_2(A)D_3(A)Q_4(A)]+\text{Tr}_x[Q_1(A)Q_2(A)D_3(A)D_4(A)]\\
&\quad+\text{Tr}_x[Q_1(A)D_2(A)Q_3(A)D_4(A)]+\text{Tr}_x[Q_1(A)D_2(A)D_3(A)Q_4(A)]\\
&\quad+\text{Tr}_x[D_1(A)Q_2(A)Q_3(A)D_4(A)]+\text{Tr}_x[D_1(A)Q_2(A)D_3(A)Q_4(A)]\\
&\quad+\text{Tr}_x[D_1(A)D_2(A)Q_3(A)Q_4(A)]-\text{Tr}_x[D_1(A)Q_2(A)Q_3(A)Q_4(A)]\\
&\quad-\text{Tr}_x[Q_1(A)D_2(A)Q_3(A)Q_4(A)]-\text{Tr}_x[Q_1(A)Q_2(A)D_3(A)Q_4(A)]\\
&\quad-\text{Tr}_x[Q_1(A)Q_2(A)Q_3(A)D_4(A)]+\text{Tr}_x[Q_1(A)Q_2(A)Q_3(A)Q_4(A)]+\circlearrowleft.
\end{split}
\end{equation}
By applying the identity (\ref{QD}) in the terms containing only one operator $Q_j(A)$, four of the six terms with two operators $Q_j(A)$ and $Q_{j'}(A)$ cancel, and we obtain
\begin{equation}
\begin{split}
T_4(A)&=\text{Tr}_x[D_1(A)D_2(A)D_3(A)D_4(A)]-\text{Tr}_x[C_{1,2}(A)D_3(A)D_4(A)]\\
&\quad-\text{Tr}_x[D_1(A)C_{2,3}(A)D_4(A)]-\text{Tr}_x[D_1(A)D_2(A)C_{3,4}(A)]\\
&\quad-\text{Tr}_x[C_{4,1}(A)D_2(A)D_3(A)]+\text{Tr}_x[Q_1(A)D_2(A)Q_3(A)D_4(A)]\\
&\quad+\text{Tr}_x[D_1(A)Q_2(A)D_3(A)Q_4(A)-\text{Tr}_x[D_1(A)Q_2(A)Q_3(A)Q_4(A)]]\\
&\quad-\text{Tr}_x[Q_1(A)D_2(A)Q_3(A)Q_4(A)]-\text{Tr}_x[Q_1(A)Q_2(A)D_3(A)Q_4(A)]\\
&\quad-\text{Tr}_x[Q_1(A)Q_2(A)Q_3(A)D_4(A)]+\text{Tr}_x[Q_1(A)Q_2(A)Q_3(A)Q_4(A)]+\circlearrowleft.
\end{split}
\end{equation}
By applying the identity (\ref{QD}) in the remaining terms containing two operators $Q_j(A)$ and $Q_{j'}(A)$, two of the four terms with three operators $Q_j(A)$, $Q_{j'}(A)$ and $Q_{j''}(A)$ cancel, and we obtain
\begin{equation}
\begin{split}
T_4(A)&=\text{Tr}_x[D_1(A)D_2(A)D_3(A)D_4(A)]-\text{Tr}_x[C_{1,2}(A)D_3(A)D_4(A)]\\
&\quad-\text{Tr}_x[D_1(A)C_{2,3}(A)D_4(A)]-\text{Tr}_x[D_1(A)D_2(A)C_{3,4}(A)]\\
&\quad-\text{Tr}_x[C_{4,1}(A)D_2(A)D_3(A)]+\text{Tr}_x[C_{1,2}(A)Q_3(A)D_4(A)]\\
&\quad+\text{Tr}_x[D_1(A)C_{2,3}(A)Q_4(A)]-\text{Tr}_x[Q_1(A)D_2(A)Q_3(A)Q_4(A)]\\
&\quad-\text{Tr}_x[Q_1(A)Q_2(A)D_3(A)Q_4(A)]+\text{Tr}_x[Q_1(A)Q_2(A)Q_3(A)Q_4(A)]+\circlearrowleft.
\end{split}
\end{equation}
Finally, by applying again the identity (\ref{QD}) in the two terms containing two operators $Q_j(A)$ and $Q_{j'}(A)$, the new terms containing three operators $Q_j(A)$, $Q_{j'}(A)$ and $Q_{j''}(A)$ combine to the remaining two terms also containing three operators $Q_j(A)$, $Q_{j'}(A)$ and $Q_{j''}(A)$, and give two terms $\text{Tr}_x[Q_1(A)Q_2(A)Q_3(A)Q_4(A)]$ with a minus sign. In conclusion, we have
\begin{equation}
\begin{split}
T_4(A)&=\text{Tr}_x[D_1(A)D_2(A)D_3(A)D_4(A)]-\text{Tr}_x[C_{1,2}(A)D_3(A)D_4(A)]\\
&\quad-\text{Tr}_x[D_1(A)C_{2,3}(A)D_4(A)]-\text{Tr}_x[D_1(A)D_2(A)C_{3,4}(A)]\\
&\quad-\text{Tr}_x[C_{4,1}(A)D_2(A)D_3(A)]+\text{Tr}_x[C_{1,2}(A)C_{3,4}(A)]\\
&\quad+\text{Tr}_x[C_{4,1}(A)C_{2,3}(A)]-\text{Tr}_x[Q_1(A)Q_2(A)Q_3(A)Q_4(A)]+\circlearrowleft.
\end{split}
\end{equation}
The trace $\text{Tr}_x[Q_1(A)Q_2(A)Q_3(A)Q_4(A)]$ can be manipulated exactly as in the case $N=3$ and we arrive to the final result
\begin{equation}
\begin{split}
T_4(A)&=\frac{1}{2}\{\text{Tr}_x[D_1(A)D_2(A)D_3(A)D_4(A)]-\text{Tr}_x[C_{1,2}(A)D_3(A)D_4(A)]\\
&\quad-\text{Tr}_x[D_1(A)C_{2,3}(A)D_4(A)]-\text{Tr}_x[D_1(A)D_2(A)C_{3,4}(A)]\\
&\quad-\text{Tr}_x[C_{4,1}(A)D_2(A)D_3(A)]+\text{Tr}_x[C_{1,2}(A)C_{3,4}(A)]\\
&\quad+\text{Tr}_x[C_{4,1}(A)C_{2,3}(A)]+\{1234\to 4321\}\},
\end{split}
\end{equation}
which again corresponds to the substitution rules given below Eq. (\ref{M_op}) for the case $N=4$.

\section{}
In this appendix, we show that the four-dimensional scalar products $(Pe_j)$ do not contain the operator $P_{\phi}$. We temporarily assume that ${k_j}^2\neq 0$ for all $j$s. In this way, by introducing the quantities $f_r^{\mu\nu}=n^{\mu}a_r^{\nu}-n^{\nu}a_r^{\mu}$, with $r=1,2$, the four-vector $e_j^{\mu}$ can be expanded with respect to the basis \cite{Baier_1976_b}
\begin{align}
\Lambda_j^{(1),\mu}&=-\frac{k_{j,\lambda}f_1^{\lambda\mu}}{k_{j,X}}, & \Lambda_j^{(2),\mu}&=-\frac{k_{j,\lambda}f_2^{\lambda\mu}}{k_{j,X}},\\
\label{Lambda_34}
\Lambda_j^{(3),\mu}&=\frac{k_j^{\mu}}{\sqrt{{k_j}^2}}, &\Lambda_j^{(4),\mu}&=-\frac{n^{\mu}{k_j}^2+k_j^{\mu}k_{j,X}}{k_{j,X}\sqrt{{k_j}^2}}
\end{align}
as $e_j^{\mu}=\sum_{u=1}^4b^{(u)}_j\Lambda_j^{(u),\mu}$, with $b^{(u)}_j=-(\Lambda_j^{(u)}e_j)$ (note that $(\Lambda_j^{(u)}\Lambda_j^{(v)})=-\delta_{uv}$, with $u,v=1,\ldots,4$). If we write the total amplitude $M$ as $M=e_{1,\mu_1}\cdots e_{N,\mu_N}M^{\mu_1\cdots\mu_N}$, then
\begin{equation}
\label{M_exp}
M=\sum_{u_1,\ldots,u_N=1}^4b^{(u_1)}_1\cdots b^{(u_N)}_N\Lambda_{1,\mu_1}^{(u_1)}\cdots\Lambda_{N,\mu_N}^{(u_N)}M^{\mu_1\cdots\mu_N}
\end{equation}
and gauge invariance requires that $k_{1,\mu_1}M^{\mu_1\cdots\mu_N}=\cdots=k_{N,\mu_N}M^{\mu_1\cdots\mu_N}=0$ \cite{Landau_b_4_1982}. This implies that the terms proportional to the four-vectors $\Lambda_{j,\mu_j}^{(3)}$ and those proportional to the divergent part of the four-vectors $\Lambda_{j,\mu_j}^{(4)}$ in the limits ${k_j}^2\to 0$, do not contribute to $M$. Thus, the amplitude $M$ remains finite in the same limits ${k_j}^2\to 0$. Moreover, the quantities $(Pe_j)$ only effectively involve contractions of $P^{\mu}$ either with $n^{\mu}$ or with $a^{\mu}_{1/2}$, so that  they do not contain the operator $P_{\phi}$. It is also worth pointing out here that in the limit ${k_j}^2\to 0$, although the contributing part of $\Lambda_{j,\mu_j}^{(4)}$ goes to zero as $\sqrt{{k_j}^2}$, the corresponding contribution to the amplitude $M$ remains finite because the quantity $b^{(4)}_j=-(\Lambda_j^{(4)}e_j)$ diverges as $1/\sqrt{{k_j}^2}$ in the same limit (see Eq. (\ref{Lambda_34})). In conclusion, by means of the above limiting procedure, our analysis can also be applied to the case in which the external photons are real, i.e., on-shell.

\bibliographystyle{elsarticle-num-names}
\bibliography{Refr_AP}

\end{document}